# Ultrafast Charge Transfer in Atomically Thin MoS$_2$/WS$_2$ Heterostructures

Xiaoping Hong*[1], Jonghwan Kim*[1], Su-Fei Shi*[1,2], Yu Zhang[3], Chenhao Jin[1], Yinghui Sun[1], Sefaattin Tongay[2,4,5], Junqiao Wu[2,4], Yanfeng Zhang[3], Feng Wang†[1,2,6]

[1] Department of Physics, University of California at Berkeley, Berkeley, CA 94720, United States

[2] Materials Science Division, Lawrence Berkeley National Laboratory, Berkeley, CA 94720, United States

[3] Department of Materials Science and Engineering, College of Engineering, Peking University, Beijing 100871, People's Republic of China

[4] Department of Materials Science and Engineering, University of California, Berkeley, CA 94720, United States

[5] School for Engineering of Matter, Transport and Energy, Arizona State University, Tempe, AZ 85287, United States

[6] Kavli Energy NanoSciences Institute at the University of California, Berkeley and the Lawrence Berkeley National Laboratory, Berkeley, California, 94720, USA

* These authors contribute equally to this work

† Correspondence to: fengwang76@berkeley.edu

**Van der Waals heterostructures have recently emerged as a new class of materials, where quantum coupling between stacked atomically thin two-dimensional (2D) layers, including graphene, hexagonal-boron nitride, and transition metal dichalcogenides ($MX_2$), give rise to fascinating new phenomena[1–10]. $MX_2$ heterostructures are particularly exciting for novel optoelectronic and photovoltaic applications, because 2D $MX_2$ monolayers can have an optical bandgap in the near-infrared to visible spectral range and exhibit extremely strong light-matter interactions[2,3,11]. Theory predicts that many stacked $MX_2$ heterostructures form type-II semiconductor heterojunctions that facilitate efficient electron-hole separation for light detection and harvesting[12–15]. Here we report the first experimental observation of ultrafast charge transfer in photo-excited $MoS_2/WS_2$ heterostructures using both photoluminescence mapping and femtosecond (fs) pump-probe spectroscopy. We show that hole transfer from the $MoS_2$ layer to the $WS_2$ layer takes place within 50 fs after optical excitation, a remarkable rate for van der Waals coupled 2D layers. Such ultrafast charge transfer in van der Waals heterostructures can enable novel 2D devices for optoelectronics and light harvesting.**

Atomically thin 2D crystals constitute a rich family of materials ranging from insulators and semiconductors to semi-metals and superconductors[1]. Heterostructures from these 2D materials offer a new platform for exploring new physics like superlattice Dirac points[4] and Hofstadter butterfly pattern[5–7], and new devices like tunnelling transistors[8], memory devices[9] and ultrathin photodetectors[2,3]. Van der Waals heterostructures of semiconducting $MX_2$ layers are particularly exciting for optoelectronic and light harvesting applications because many $MX_2$ monolayers are direct-bandgap semiconductors[16,17] with remarkably strong light-matter interactions[2,3,11]. And importantly, $MX_2$ heterostructures are predicted to form type-II heterojunctions that can assist efficient separation of photoexcited electrons and holes[12–15].

In type-II heterojunctions, the conduction band minimum and valence band maximum reside in two separate materials. Photoexcited electrons and holes therefore prefer to stay at separate locations. Figure 1a illustrates the alignment of electronic bands of $MoS_2$ and $WS_2$ monolayers predicted by a recent theory[12]. It shows that monolayer $MoS_2$ and $WS_2$ has a bandgap of 2.39 eV and 2.31eV, respectively, and the $MoS_2$ valence band maximum is 350 meV lower than that of $WS_2$. Consequently, the $MoS_2/WS_2$ heterostructure forms a type-II heterojunction (if we neglect the hybridization of electronic states in $MoS_2$ and $WS_2$ layers), with the conduction band minimum residing in $MoS_2$ and the valence band maximum in $WS_2$, respectively. In the single-particle picture, this heterojunction will lead to efficient charge transfer with separated electron and holes in two layers upon optical excitation (Figure 1a), which can have a dominating effect on both light emission and photovoltaic responses in $MoS_2/WS_2$ heterostructures.

However, there are two outstanding questions regarding charge transfer processes in the atomically thin and van der Waals-coupled $MoS_2/WS_2$ heterostructure. (1) How do strong electron-electron interactions and excitonic effects affect charge transfer processes? (2) How fast can charge transfer take place between van der Waals-coupled layers? Electron-electron interactions are dramatically enhanced in 2D materials due both to size confinement and inefficient screening. Theoretical studies[18,19] have predicted an exciton binding energy from 500 meV to 1 eV in $MX_2$ monolayers, which is larger than the expected band displacement of 350 meV in $MoS_2/WS_2$ heterostructure. Therefore the exciton cannot dissociate into a free electron and a free hole in two separate layers. Will this large exciton binding energy then prevent charge transfer processes and keep the exciton in one layer, or will a new bound state of layer-separated electron and hole pair be generated? In addition, van der Waals coupling is rather weak compared to covalent bonding. Will that lead to a much slower charge transfer process in van der Waals heterostructures compared to their covalent counterparts? In this letter, we address these questions experimentally. Through combined photoluminescence spectroscopy and optical pump-probe spectroscopy, we demonstrate that ultrafast charge transfer takes place very efficiently in $MoS_2/WS_2$ heterostructures. In particular, holes in the $MoS_2$ layer can separate into the $WS_2$ layer within 50 fs upon photo-excitation.

Figure 1b schematically shows our sample configuration. In brief, $MoS_2$ monolayers were grown on 285 nm $SiO_2$/Si substrates using the chemical vapour deposition (CVD) method[20]. They were subsequently transferred on top of as-grown CVD $WS_2$ flakes on sapphire substrates[21] to form $MoS_2/WS_2$ heterostructures. Raman spectra (Figure 1c) from isolated $MoS_2$ and $WS_2$ films confirm that both are monolayers because the energy separation between Raman active modes agrees well with previous reported values for monolayer $MoS_2$ and $WS_2$.[22,23] The Raman spectrum of a $MoS_2/WS_2$ heterostructure (Figure 1c) appears to be an addition of Raman modes from the constituent layers.

One sensitive probe of charge transfer in $MX_2$ heterostructures is photoluminescence (PL) spectroscopy, because an electron and hole pair spatially separated in two $MX_2$ layers cannot emit efficiently. We performed PL spectroscopy and mapping on multiple $MoS_2/WS_2$ heterostructure samples. Figure 2a displays the optical image of one sample where a large continuous $MoS_2$ piece (covering everywhere in the image) was transferred on top of $WS_2$ flakes (the bright areas). Figure 2b shows the PL intensity map at the $MoS_2$ A-exciton resonance (1.93 eV) at 77 K when the sample is excited by 2.33 eV photons. We observed strong PL signals in the $MoS_2$-only region, but the PL is significantly quenched in the $MoS_2/WS_2$ heterostructure region. Figure 2c further displays typical PL spectra for $MoS_2/WS_2$ heterostructures, isolated $MoS_2$, and isolated $WS_2$ layers with 2.33 eV excitation. It is apparent that $MoS_2$ and $WS_2$ monolayers show strong PL at their respective A-exciton resonances (1.93 eV and 2.06 eV), but both PL signals are efficiently quenched in $MoS_2/WS_2$ heterostructures. Room temperature PL spectra also exhibit similar behaviour. (See Supplementary Information Part 1). In principle, PL signals can be quenched by two mechanisms in a heterostructure: energy transfer and charge transfer. However, energy transfer quenches only the PL from a higher energy transition (i.e. 2.06 eV resonance in $WS_2$), but tends to enhance luminescence from the lower energy transition (i.e. 1.93 eV resonance in $MoS_2$). On the other hand, charge transfer will quench light emission from all transitions. Therefore the observation of reduced PL from both $WS_2$ and $MoS_2$ exciton resonances in $MoS_2/WS_2$ heterostructures demonstrates that efficient charge transfer takes place in this type-II heterojunction.

To directly probe the charge transfer process and its ultrafast dynamics, we measured transient absorption spectra of $MoS_2/WS_2$ heterostructures using resonant pump-probe spectroscopy. A femtosecond pulse first excites the heterostructure, and the photo-induced changes in the reflection spectrum ($\Delta R/R$) are probed by a laser-generated supercontinuum light after controlled time delays. For atomically thin heterostructures on a transparent sapphire substrate, the reflection change $\Delta R/R$ is directly proportional to the change in absorption coefficient[24]. $MoS_2$ and $WS_2$ monolayers have distinctly different exciton transitions. Therefore we can selectively excite the $MoS_2$ or $WS_2$ layer using specific resonant optical excitations, and probe the accumulation of electrons and holes in different layers through photo-induced changes in their respective exciton transitions. Specifically, we choose a pump photon energy at 1.86 eV to excite exclusively the A-exciton transition of $MoS_2$. This pump cannot excite $WS_2$ directly because the photon energy is far below the absorption threshold of $WS_2$. Afterwards we examine the photoinduced changes of both $WS_2$ and $MoS_2$ exciton resonances in transient absorption spectra from 2.0-2.5 eV to probe the charge distribution in heterostructures.

Using a pump fluence of 85 μJ/cm$^2$, A-excitons in $MoS_2$ with a density $\sim 5\times10^{12}/cm^2$ are generated immediately after photo-excitation. Figure 3a shows a two-dimensional plot of transient absorption spectra in a $MoS_2/WS_2$ heterostructure at 77 K, where the colour scale, the horizontal axis, and the vertical axis represent the magnitude of -$\Delta R/R$, the probe photon energy, and pump-probe time delay, respectively. The figure shows prominent resonant features in transient absorption centred on 2.06 eV and 2.46 eV, with the higher energy feature several times weaker than the lower energy one. Comparing with linear absorption spectra of isolated $WS_2$ and $MoS_2$ monolayers (Figure 3e), we can attribute these two resonant features respectively to the A and B-exciton transitions in $WS_2$, although the $WS_2$ layer is not excited by the pump. To better understand the transient absorption spectra in $MoS_2/WS_2$ heterostructures, we also performed control experiments for isolated $WS_2$ and $MoS_2$ monolayers. In bare $WS_2$ monolayers no pump-induced signal can be observed above the noise level, consistent with the fact that no direct absorption can take place in $WS_2$. (Supplementary Information Part 2). In isolated $MoS_2$ monolayers, pump-induced absorption changes in our spectral range is centred at 2.11 eV (Figure 3b), corresponding to the B-exciton transition of $MoS_2$. Figure 3c-d show detailed comparisons of transient absorption spectra in a $MoS_2/WS_2$ heterostructure and an isolated $MoS_2$ monolayer at the pump-probe time delay of 1 picosecond (ps) (Figure 3c) and 20 ps (Figure 3d). Although the resonant features at 2.06 eV for the heterostructure and at 2.11 eV for monolayer $MoS_2$ are close in energy, they are clearly distinguishable and match well with the A-exciton in $WS_2$ and B-exciton in $MoS_2$ in the absorption spectra (Figure 3e), respectively. In addition, the transient absorption signal at the $WS_2$ A-exciton transition in the heterostructure is stronger in magnitude and has a narrower spectral width and a slower decay time constant.

Our transient absorption measurements of $MoS_2/WS_2$ heterostructures establish unambiguously that optical excitation in $MoS_2$ leads to strong modification of exciton transitions in $WS_2$, which has a larger optical bandgap. It provides direct evidence of efficient charge separation in photo-excited $MoS_2/WS_2$ heterostructures as described in Figure 1a: electron hole pairs are initially created in the $MoS_2$ layer, but holes quickly transfer to the WS$_2$ layer due to the type-II band alignment, while electrons stay in the MoS$_2$ layer. The photo-excited electrons in MoS$_2$ and holes in WS$_2$ lead to strong transient absorption signal for exciton transitions in both MoS$_2$ and WS$_2$. Transient absorption signals are the strongest for the A-excitons due to their sharper resonances and efficient photo-bleaching effects from Pauli blocking, but B-exciton transitions are also affected. Consequently, the transient absorption spectra in MoS$_2$/WS$_2$ heterostructures are dominated by the A-exciton transition in WS$_2$. (The A-exciton transition of MoS$_2$ is outside of the probe spectral range because it overlaps with the pump wavelength). Photo-induced changes of B-exciton transitions in the MoS$_2$/WS$_2$ heterostructure (Figure 3a) and in the MoS$_2$ monolayer (Figure 3b) can also be identified, but they are significantly weaker than that of A-exciton transitions. Room temperature data show similar trends and are included in the Supplementary Information Part 3.

The rise time of the WS$_2$ A-exciton transient absorption signal probes directly the hole transfer dynamics from the MoS$_2$ layer because this signal exists only after the hole transfer, but not right at the excitation of MoS$_2$. Figure 4 shows the dynamic evolution of the WS$_2$ A-exciton resonance in the MoS$_2$/WS$_2$ heterostructure (Figure 4a), which can be compared to the transient absorption signal for the B-exciton resonance in an isolated MoS$_2$ monolayer (Figure 4b). We found that the rise time in both signals are almost identical, and it is limited by the laser pulse duration ~ 250 fs. In Figure 4b the MoS$_2$ monolayer is directly pumped, and the photo-induced signal should appear instantaneously. We could reproduce the ultrafast dynamics in the MoS2 monolayer in Figure 4b by convoluting the instrument response function (the blue dashed curve in Figure 4b) with an instantaneous response in MoS2. Using the same instrument response for time convolution, we can then reproduce the experimentally observed signal in the heterostructure with a rise time shorter than 50 fs (red line in Figure 4a). Therefore our results show that holes are transferred from the $MoS_2$ layer to the $WS_2$ layer within 50 fs after optical excitation of the $MoS_2/WS_2$ heterostructure, a remarkably fast rate. Similar ultrafast hole transfer also happens at room temperature, as shown in Supplementary Information Part 4. This hole transfer time is much shorter than the exciton lifetime and most other dynamics processes, which are on the order of several to tens of picosecond[25]. Electrons and holes can therefore be efficiently separated into different layers immediately after their generation. Consequently, PL from $MoS_2$ and $WS_2$ exciton resonances will be strongly quenched, as we observed previously.

Our experimental data establishes that charge separation in $MoS_2/WS_2$ heterostructures is very efficient, although the band offset between $MoS_2$ and $WS_2$ is smaller than the predicted exciton binding energy in monolayer $MX_2$. Energetically uncorrelated free electrons and holes in separated $MoS_2$ and $WS_2$ layers cannot be produced through the excitation of $MoS_2$ A-excitons. However, the $MoS_2$ and $WS_2$ layers are only ~0.62 nm separated from each other[13], suggesting that even for layer-separated electrons and holes, strong Coulomb interactions can lead to bound exciton states. These exciton states with electron and hole residing in different layers can be energetically favourable compared to an exciton confined to only $MoS_2$ layer, and are likely to be responsible for the efficient charge separation observed in $MoS_2/WS_2$ heterostructures. Such bounded excitons with electron and hole in different materials, known as charge transfer excitons (CTC)[26,27], have also been investigated in other type-II heterojunctions, such as molecule/nanocrystals and organic/inorganic interfaces[26–28].

The observed sub-50 fs hole transfer time is remarkably short considering that the $MoS_2$ and $WS_2$ layers are randomly stacked and are coupled by relatively weak van der Waals interactions. In comparison, the charge transfer time in conventional covalently-bonded type-II heterostructures ranges from sub-100 fs to 100 ps depending on specific electronic couplings and band alignments[28]. One factor contributing to the ultrafast charge transfer rate in atomically thin heterostructures is the close proximity of the two heterolayers, because electrons or holes only need to move less than 1nm vertically for the charge transfer process to happen. Still, the 50 fs hole transfer time for van der Waals heterostructures is surprisingly fast. Microscopic understanding of this ultrafast hole transfer in $MX_2$ heterostructures requires detailed theoretical studies to examine the hybridization of electronic states in twisted heterolayers and the dynamic evolution of photo-excited states due to electron-phonon and electron-electron interactions. It is known that for $MoS_2$ bilayers, electronic coupling at the K point in the Brillouin zone is rather weak. Electron wavefunction hybridization at the $\Gamma$ point, however, is much stronger, which leads to a rise in $\Gamma$ point valence band and an indirect bandgap in bilayer $MoS_2$. Electronic coupling between incommensurate $MoS_2$ and $WS_2$ will play an important role in the charge transfer dynamics of twisted $MoS_2/WS_2$ heterostructures, the behaviour of which has been little studied so far. Because van der Waals heterostructures have atomically sharp interfaces with no dangling bonds and well-defined optical resonances, they provide an ideal model system for further experimental and theoretical investigations of interfacial charge transfer processes and the charge transfer exciton states.

The ultrafast charge transfer process in atomically thin $MX_2$ heterostructures has important implications for photonic and optoelectronic applications. $MX_2$ semiconductors have extremely strong optical absorption, and has been considered previously for photodetectors[2,3,11], photovoltaics[29] and photocatalysis[30]. However, the large exciton binding energy in $MX_2$ thin films poses a challenge for efficient separation of photo-generated electron and holes. The type-II $MX_2$ heterostructures, with femtosecond charge transfer rate, provides an ideal way to spatially separate electrons and holes for electrical collection and utilization.

In summary, we have demonstrated for the first time efficient charge transfer in $MoS_2/WS_2$ heterostructures through combined PL mapping and transient absorption measurement. We quantitatively determine the ultrafast hole transfer time to be less than 50 fs, a surprisingly fast rate for van der Waals coupled materials. Our study suggests that $MX_2$ heterostructures, with their remarkable electrical and optical properties and rapid development of large area synthesis, hold great promise for future optoelectronic and photovoltaic applications.

**Methods**

**$MX_2$ monolayer Growth**:

$MoS_2$: Monolayer $MoS_2$ was grown by CVD on 285 nm $SiO_2$/Si substrates[20]. Substrates were loaded into a 1-inch CVD furnace and placed face-down above a ceramic boat containing 4.2 mg of $MoO_3$ (≥99.5% Sigma-Aldrich). A crucible containing 150 mg of sulfur (≥99.5% Sigma-Aldrich) is placed upstream. The CVD growth is performed at atmospheric pressure with flowing ultrahigh-purity nitrogen. Tuning the sulfur concentration can roughly modify the nucleation density and control the transition of triangular single crystals to large area monolayer.

WS$_2$: Large area WS2 monolayer is grown on sapphire substrates by CVD[21]. A multi-temperature-zone tube furnace (Lindberg/Blue M) equipped with a 1-inch-diameter quartz tube was used for the growth. Sulfur powder was mildly sublimated at ~100°C and placed outside the hot zone. WO3 powder (Alfa Aesar, purity 99.9%) and sapphire substrates (<0001> oriented single crystals) were successively placed in the hot centre. We used Argon (flow rate 80 sccm) (or mixed Ar and H2 gas with flow rate of 80 and 10 sccm, respectively) to carry WO3-x vapour species to the downstream substrates. The growth pressure was set at 30 Pa. Growth temperature is set at ~900°C and growth time is set at ~60 min.

**Heterostructure preparation**: The heterostructure was prepared by transferring[20] monolayer MoS$_2$ onto monolayer WS$_2$ on sapphire. The CVD grown MoS$_2$ single layer (described above) on SiO2/Si is spin-coated with PMMA (A4) at 4000 rpm for 60 seconds. PMMA/ MoS$_2$ film is separated from the substrate (SiO2/Si) by KOH etching (1 mol/L) at 80 °C. The film is transferred to DI water beakers to dilute KOH residue under MoS$_2$. Finally, it is transferred on to CVD grown WS$_2$ on sapphire substrate (described above) and soaked in acetone to dissolve PMMA.

**Photoluminescence measurements**: We use a 532 nm laser (photon energy =2.33 eV) to excite the isolated monolayers of MoS$_2$, WS$_2$, and MoS$_2$/WS$_2$ heterostructures. The laser beam is focused to a diffraction-limited spot with a diameter ~ 1 μm, and the PL is collected in the reflection geometry via confocal microscope. A monochrometer and a liquid nitrogen cooled CCD are used to record the PL spectra. Two-dimensional PL mapping is done by scanning the computer-controlled piezoelectric stage.

**Linear absorption spectra**: A supercontinuum laser (Fianium SC450) is used as broadband light source. The laser is focused at the sample with ~2 μm beam size and the reflection signal R is collected via confocal microscopy and analysed by a spectrometer equipped with a one-dimensional CCD array. Reference spectrum $R_0$ is taken on the sapphire substrate nearby the sample (isolated $MoS_2$, isolated $WS_2$ and heterostructure). The normalized difference signal $(R-R_0)/R_0$ is directly proportional to the linear absorption from atomically thin layers on sapphire.

**Pump-probe measurement:** Femtosecond pulses at 1026 nm are generated by a regenerative amplifier seeded by a mode-locked oscillator (Light Conversion PHAROS). The femtosecond pulses (at a repetition rate of 150 kHz and a pulse duration ~250 fs) are split into two parts. One part is used to pump an optical parametric amplifier to generate tunable excitation laser pulses, and the other part is focused into a sapphire crystal to generate a supercontinuum white light (500 nm ~ 900 nm) for probe pulses. Pump and probe beam are focused at the sample with diameters about 50 μm and 25 μm, respectively. The probe light is detected by a high-sensitivity photomultiplier after wavelength selection through a monochromater with spectral resolution of 1 nm. The pump-probe time delay is controlled by a motorized delay stage, and the pump-probe signal is recorded using the lock-in detection with a chopping frequency of 1.6 kHz.

**Figure captions**

**Figure 1: Band alignment and structure of MoS$_2$/WS$_2$ heterostructures. a.** Schematics of theoretically predicted band alignment of a MoS$_2$/WS$_2$ heterostructure, which forms a type-II heterojunction. Optical excitation of the MoS$_2$ A-exciton will lead to layer separated electron and hole carriers. **b.** Illustration of a MoS$_2$/WS$_2$ heterostructure with a MoS$_2$ monolayer lying on top of a WS$_2$ monolayer. Electrons and holes created by light are shown to separate into different layers. **c.** Raman spectra of an isolated MoS$_2$ monolayer (blue trace), an isolated WS$_2$ monolayer (red trace), and a MoS$_2$/WS$_2$ heterostructure (black trace).

**Figure 2 Photoluminescence (PL) spectra and mapping of MoS$_2$/WS$_2$ heterostructures at 77 K. a.** Optical microscope image of a typical MoS$_2$/WS$_2$ heterostructure sample. The MoS$_2$ layer covers everywhere in the image, and bright areas correspond to MoS$_2$/WS$_2$ heterostructures. **b.** PL mapping data taken in the dashed rectangle area in **a**. The colour scale represents PL intensity at the MoS$_2$ A-exciton resonance (1.93 eV). It clearly shows that MoS$_2$ PL is strongly quenched in the heterostructure. Scale bar is 5 μm. **c.** Typical PL spectra of an isolated monolayer MoS$_2$, an isolated monolayer WS$_2$ and a MoS$_2$/WS$_2$ heterostructure. The isolated MoS$_2$ and WS$_2$ monolayer show strong PL at 1.93 eV and 2.06 eV, respectively, corresponding to their A-exciton resonances. Both exciton PL signals are strongly quenched in the MoS$_2$/WS$_2$ heterostructure, suggesting an efficient charge transfer process exists in the heterostructure.

**Figure 3 Transient absorption spectra of MoS$_2$/WS$_2$ heterostructures. a.** and **b.** Two-dimensional plots of transient absorption spectra at 77 K from a MoS$_2$/WS$_2$ heterostructure (**a**) and an isolated MoS$_2$ monolayer (**b**) upon excitation of the MoS$_2$ A-exciton transitions. The horizontal axis, vertical axis, and colour scale represent the probe photon energy, pump-probe time delay, and the transient absorption signal, respectively. Positive signals indicate a pump-induced decrease in absorption. **c.** and **d.** Transient absorption spectra for MoS$_2$/WS$_2$ (red circles) and MoS$_2$ (green square) at 1 ps and 20 ps pump-probe delay, respectively. **e.** Linear absorption spectra of monolayers of MoS$_2$ (magenta line) and WS$_2$ (blue line). Although only MoS$_2$ A-exciton transitions are optically excited, transient absorption spectra in the MoS$_2$/WS$_2$ heterostructure are dominated by a resonance feature (red circles in c and d) corresponding to the WS$_2$ A-exciton transition (blue line in e), which is clearly distinguishable from the resonance feature corresponding to the MoS$_2$ B-exciton transition in an isolated MoS$_2$ monolayer (green squares in c and d and magenta line in e). It demonstrates unambiguously an efficient hole transfer from the photoexcited MoS$_2$ layer to the WS$_2$ layer in MoS$_2$/WS$_2$ heterostructures.

**Figure 4 Ultrafast hole transfer dynamics from vertical cuts in Figure 3a and b. a.** The evolution of transient absorption signals at the WS$_2$ A-exciton resonance in the MoS$_2$/WS$_2$ heterostructure. **b.** The dynamic evolution of transient absorption signals at the MoS$_2$ B-exciton resonance in the isolated MoS$_2$ monolayer. Both signals show almost identical ultrafast rise time, which is limited by the laser pulse duration ~ 250 fs. By convoluting the instrument response function (blue dashed line in b) and an instantaneous response in MoS$_2$, we can reproduce the ultrafast dynamics in the MoS$_2$ monolayer (red

trace in b). Similar convolution shows that the rise time in MoS$_2$/WS$_2$ monolayer is around 25 fs (red trace in a), and has an upper limit of 50 fs. It demonstrates that holes can transfer from the photoexcited MoS$_2$ layer to the WS$_2$ layer within 50 fs in the MoS$_2$/WS$_2$ heterostructure.

**Acknowledgement**: Optical measurements and MoS$_2$ growth of the work were supported by Office of Basic Energy Science, Department of Energy under contract No. DE-SC0003949 (Early Career Award) and No. DE-AC02-05CH11231 (Materials Science Division). The WS$_2$ growth part was financially supported by the National Natural Science Foundation of China (Grants Nos. 51222201, 51290272) and the Ministry of Science and Technology of China (Grants No. 2011CB921903). F.W. acknowledges the support from a David and Lucile Packard fellowship.

**Author contributions**: F.W. conceived and supervised the experiment. X.H., J.K. and S.-F.S. carried out PL and pump probe measurements. Y.S., S.T. and J.W. grew CVD monolayer MoS$_2$. Y.Z and Y.F.Z. grew CVD monolayer WS$_2$. J.K. X.H. and S.-F.S prepared heterostructure sample. X.H., J.K., S.-F.S., C.J. performed data analysis. All authors discussed the results and wrote the manuscript.

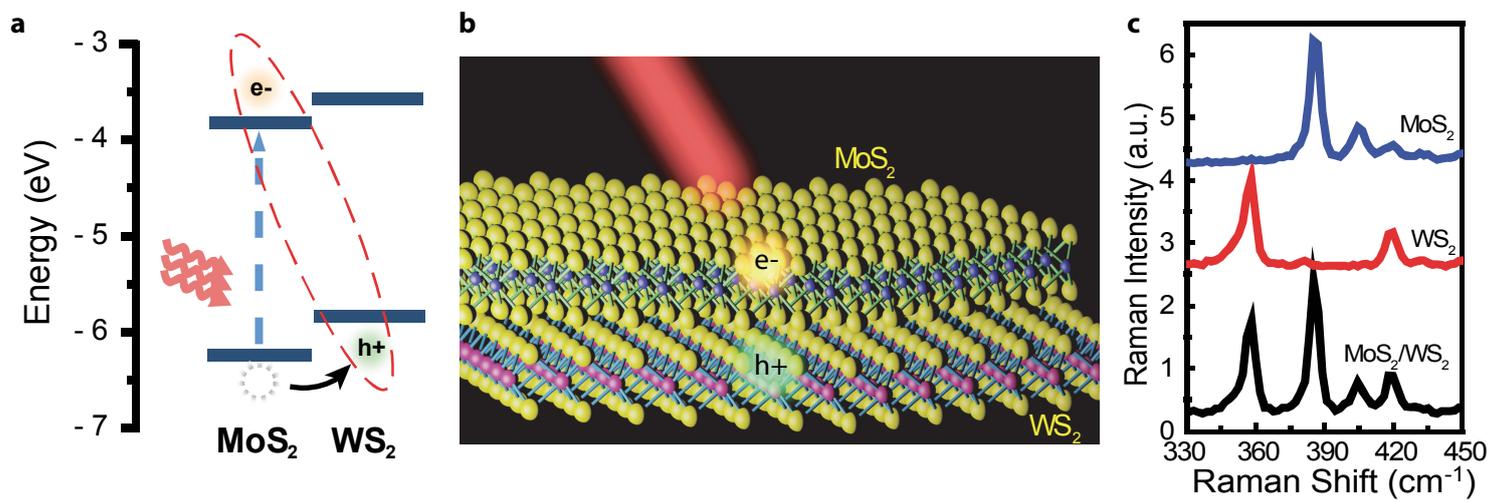

Figure 1

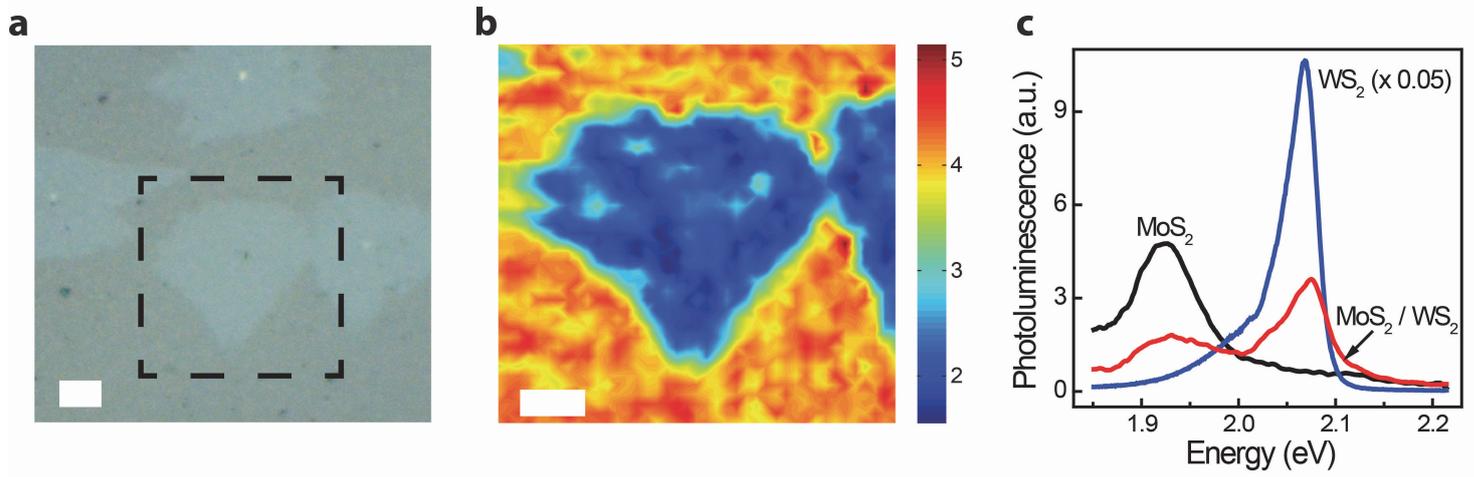

Figure 2

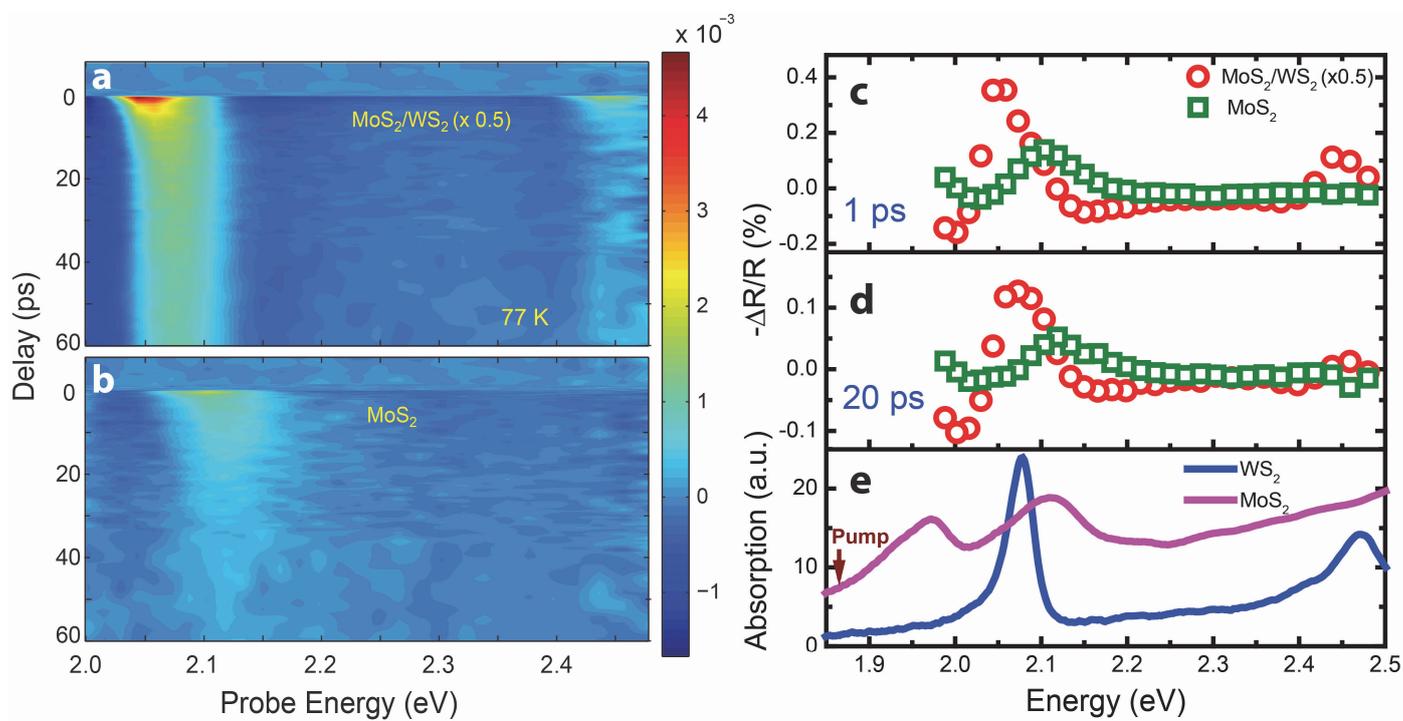

Figure 3

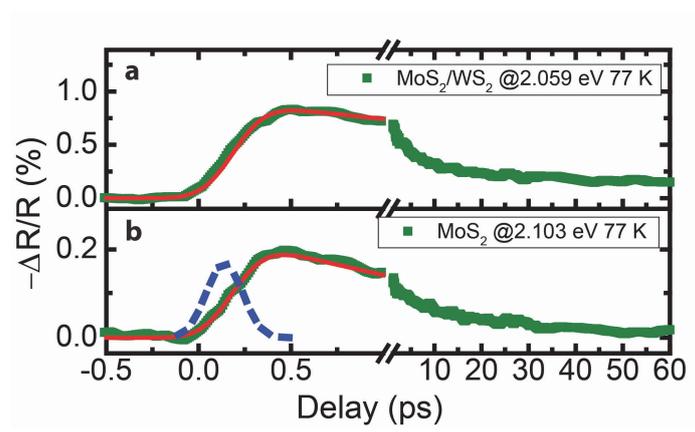

Figure 4